\documentclass[twocolumn,showpacs,preprintnumbers,amsmath,amssymb,floatfix,prstab]{revtex4}

\usepackage{graphicx}
\usepackage{subfigure}
\usepackage{dcolumn}
\usepackage{bm}
\usepackage{ctable}
\usepackage{multirow}

\begin{document}
\title{A New Item Response Theory Model for Open-Ended Online Homework with Multiple Allowed Attempts}
\author{Emre G\"on\"ulate\c{s}}
\email{gonulat1@msu.edu}
\affiliation{%
Department of Educational Psychology and Special Education,
Michigan State University, East Lansing, MI 48824, USA
}%
\author{Gerd Kortemeyer}
\email{kortemey@msu.edu}
\affiliation{%
Lyman Briggs College and Department of Physics and Astronomy,
Michigan State University, East Lansing, MI 48824, USA
}%
\date{\today}
\begin{abstract}
Item Response Theory (IRT) was originally developed in traditional exam settings, and it has been shown that the model does not readily transfer to formative assessment in the form of online homework. We investigate if this is mostly due to learner traits that do not become apparent in exam settings, namely random guessing due to lack of diligence or dedication, and copying work from other students or resources. Both of these traits mask the true ability of the learner, which is the only trait considered in most mainstream unidimensional IRT models. We find that indeed the introduction of these traits allows to better assess the true ability of the learners, as well as to better gauge the quality of assessment items. Correspondence of the model traits to self-reported behavior is investigated and confirmed. We find that of these two traits, copying answers has a larger influence on initial homework attempts than random guessing.
\end{abstract}

\maketitle

\section{Introduction}
Item Response Theory (IRT)  is gaining increased attention in discipline-based educational research, as it models the interplay between learner traits and assessment item properties (the word ``item'' in this context denotes what physics educators would call a ``problem''). As opposed to Classical Test Theory, IRT assumes that learners have latent traits beyond their overt score on a particular set of test items; the same  homework, practice, concept inventory, or exam problem may ``work'' differently for different learners. By the reverse token, the same learner may have different scores on different sets of test items, depending on how difficult, well-written, meaningful, or representative these items are. In recent years, within Physics Education Research, IRT has been used to examine the validity of concept tests (e.g.,~\cite{lin09,cardamone11}) and online homework (e.g.~\cite{lee08,kortemeyer14a}).

IRT was originally developed in traditional exam settings, which are highly controlled and allow only one attempt to arrive at the correct solution. The same theory does not easily transfer to online homework, which typically gets completed in open environments (e.g., at home with both access to study resources and distractions, or in libraries and study lounges with interaction among learners) and allows for multiple attempts.

The performance on a particular attempt at solving homework may not necessarily be a true reflection of the learner's ability; most notably, noise is introduced to the data through guessing (some students trying out some random solutions or educated guesses~\cite{kortemeyer09}) and copying (some students copying or extensively collaborating on homework solutions~\cite{kortemeyer05ana,kortemeyer07correl,palazzo10}). 

Most mainstream unidimensional IRT models only consider one learner trait, usually called ``ability.'' This assumes that the probability of success in solving a homework problem depends mostly on how capable the learner is, which in turn is likely a mixture of knowledge, practice, intelligence, and general problem-solving ability. Results in applying this one-trait model to online homework have been encouraging~\cite{kortemeyer14a}, but there are notable discrepancies between the ability traits derived from homework data and those derived from exam data.

This study proposes to model and absorb some of the noise by introducing additional learner traits beyond ability, namely a particular learner's likelihood to randomly guess on an item (thus under-representing their true ability), or to copy from another learner (thus over-representing their own true ability).  There are good indicators that these traits vary between students: some students copy more than others~\cite{palazzo10}, and some students are more careless in guessing solutions than others (for example, male learners are more prone to try out ``random solutions than female learners, and they also spend less time to reconsider a problem between subsequent attempts~\cite{kortemeyer09}).

Sect.~\ref{sec:prior} provides relevant results from prior research, Sect.~\ref{sec:new} introduces the new model, Sect.~\ref{sec:results} presents results from running the full and reduced models on data from first attempts on homework, Sect.~\ref{sec:disc} discusses these results, Sect.~\ref{sec:outlook} gives and outlook on future opportunities and challenges, and Sect.~\ref{sec:conclusions} concludes the paper.

\section{Prior Results}\label{sec:prior}
An earlier study of extending IRT to formative assessment was carried out in a large enrollment (256 student) physics course for scientists and engineers~\cite{kortemeyer14a}. The course has a large number of exams throughout the semester (12 midterm quizzes and one final), which resulted in 184 exam items, and it has 13 online homework assignments with a total of 401 problems. Item properties and a latent learner ability trait were modeled using a standard  two parameter logistic (2PL) model~\cite{birnbaum}, in which the probability of learner $j$ solving problem $i$ is modeled as
\begin{equation}\label{eq:2PL}
p_{ij}=\frac{1}{1+\exp\left(a_i(b_i-\theta_j)\right)}\ .
\end{equation}
Here, $\theta_j$ models the ability of learner $j$, $b_i$ the difficulty of item $i$, and $a_i$ the discrimination of item $i$. IRT estimation algorithms determine these parameters and traits by a coupled iterative optimization process, in which the values are adjusted in each step to better fit the actual assessment outcome.

\begin{figure}
\begin{center}
\includegraphics[width=0.49\textwidth]{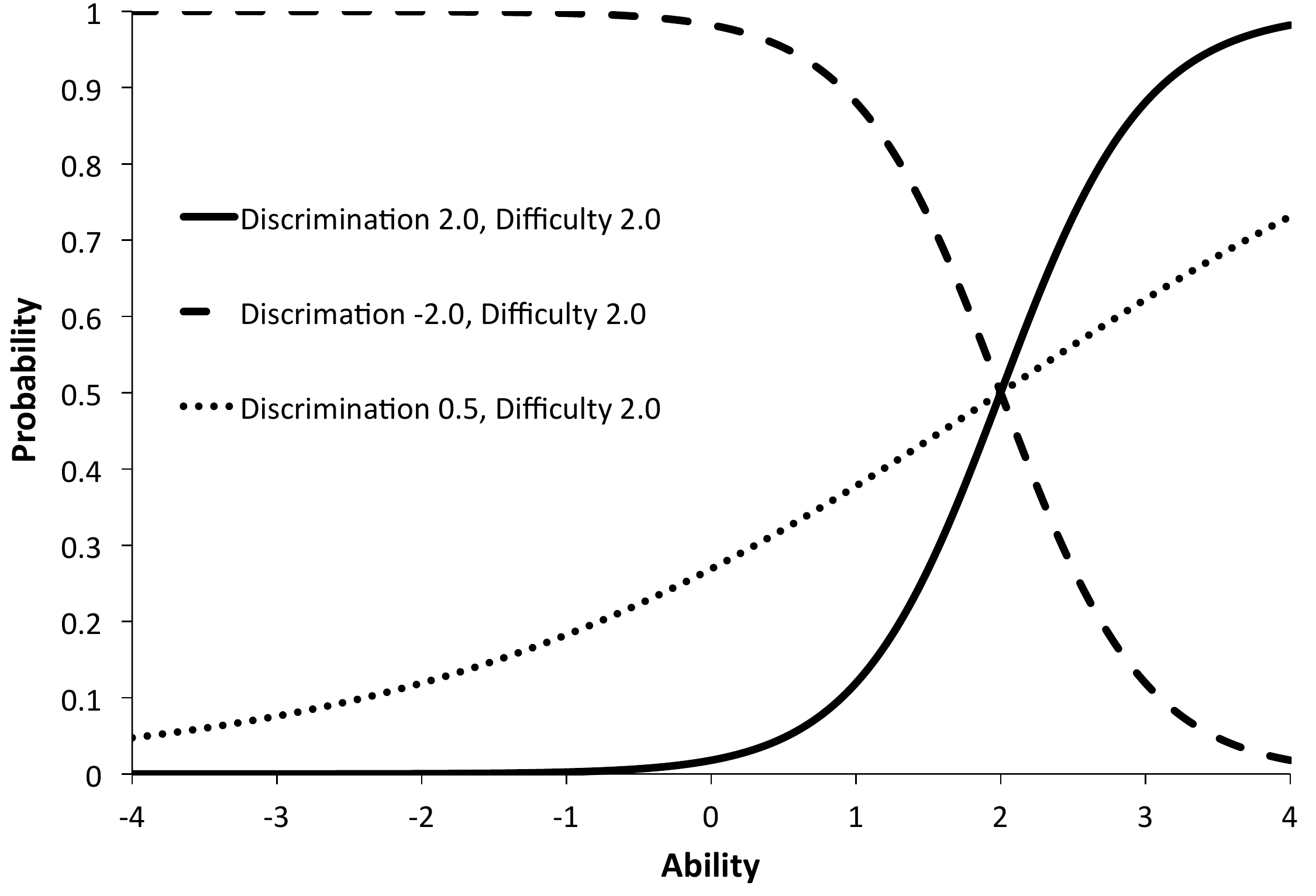}
\end{center}
\caption{Examples of item characteristic curves for different discrimination and difficulty parameters~\cite{kortemeyer15b}. The abscissa is student ability $\theta_j$, the  ordinate the function $p_{ij}=p_i(\theta_j)$ for different $a_i$ and $b_i$.}
\label{fig:itemcharacteristic}
\end{figure}

What each of the parameters does can best be illustrated using the graph of the function $p_{ij}$, which is known as the item characteristic curve. Fig.~\ref{fig:itemcharacteristic} shows examples of item characteristic curves with different values of $a_i$ and $b_i$. For an item with positive discrimination, a high-ability student is more likely to solve it than a low-ability student. How rapidly the probability changes with increasing ability is determined by the discrimination parameter $a_i$, which determines the slope at the point of inflection that is determined by the difficulty $b_i$. This difficulty parameter shifts the whole curve to the left or the right.

This functional form is somewhat arbitrary: essentially, Eq.~\ref{eq:2PL} just happens to be a function with the right asymptotic properties and a transition between likely-to-not-solve and likely-to-solve that can be controlled easily by a small number of meaningful parameters. Similar models could have been built by for example using a parametrized arctangent or hyperbolic tangent, as long as the asymptotic values are zero for infinitely
low-ability and unity for infinitely high-ability learners, however, Eq.~\ref{eq:2PL} is the traditional and most straightforward implementation.

Good formative assessment problems have medium difficulty: they are not too hard, so they do not frustrate the majority of learners, but they are also not so easy to be meaningless. They also have high positive discrimination, so they give meaningful feedback to both learners and instructors. An item with negative discrimination is unusable: low-ability students have a better chance of solving it than high-ability students (maybe due to a subtle difficulty that lower ability students overlook, or simply due to an error).

\begin{figure*}
\begin{center}
\includegraphics[width=0.49\textwidth]{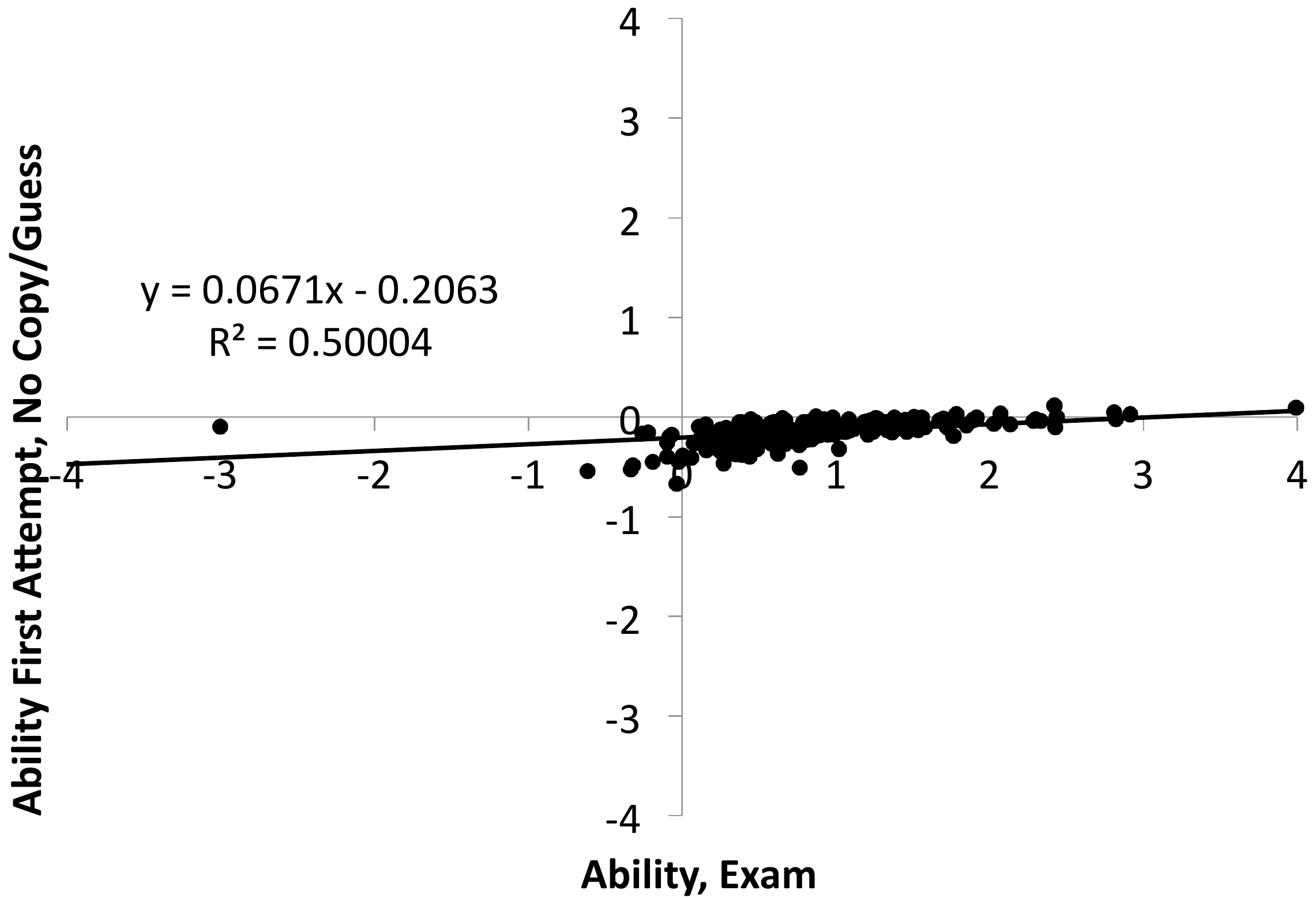}
\includegraphics[width=0.49\textwidth]{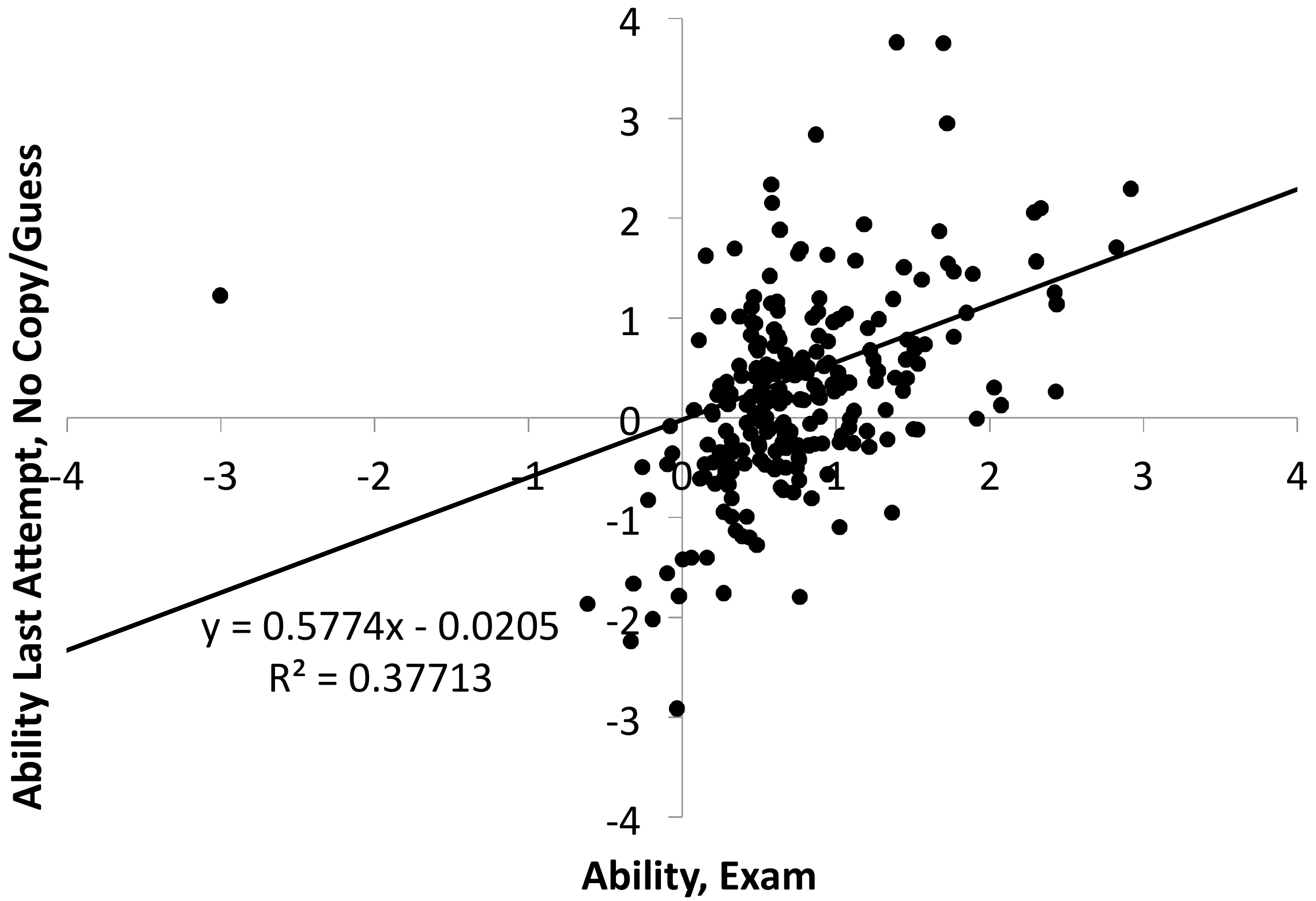}
\end{center}
\caption{Relationship between the student abilities $\theta_i$ (Eq.~\ref{eq:2PL}) derived from homework and exam data. East data point represents one student in the course; linear fits are presented to assess correlations. The left panel shows ability based on success on the very first attempt, while the right panel is based on eventual success (i.e., success on the last attempt).}
\label{fig:firstlastnone}
\end{figure*}

When using IRT for online homework with multiple attempts, frequently the initial (first) attempt that the learner made on the homework is considered. This choice is very reasonable: after all, on exams, learners only have one attempt to arrive at the correct solution. However, assuming that exams are reliable indicators of student ability, it was found that the data from the first attempt on homework is no better indicator of student ability than the data from the eventual homework outcome (after the final (last) attempt)~\cite{kortemeyer14a}. In fact, using the same data, a simple correlation between abilities obtained from the initial and final attempts on the one hand and exam data on the other shows that overall the final attempt may be a better indicator in a 2PL model, see Fig.~\ref{fig:firstlastnone}: the first attempt may explain more of the variation, but it has less variation. If no other learner traits are considered, this result may be interpreted as initial genius having less impact on physics ability (as measured by exams) than tenacity to eventually solve the problem.

While the tenacity argument is certainly convincing and even somewhat encouraging, it leaves a lot of the noise unexplained: why do low-ability students get some homework correct on the first attempt, and why do even high-ability students fail on some homework? The answer is likely that students know the rules of the game in homework: they know that they are allowed to guess and guessing will not hurt their scores, and they know that copying and collaboration will go largely undetected.

As guessing and copying are clearly present in online homework~\cite{palazzo10,kortemeyer15a}, can some of the noise in the data be absorbed into additional parameters? 
In an earlier study, additional {\it item} parameters were introduced that would allow to model the guessing and copying of answers on particular homework problems~\cite{kortemeyer14a}. This, however, failed to bring about any improvements, which in retrospect is not surprising: the amount of guessing and copying is likely not a property of a particular homework problem, but rather a trait of the learner. In other words, if a particular student guesses or copies while working on a particular homework problem does not depend on the problem but on the learner. Can the predictive power of homework be increased by taking into consideration these behaviors by introducing additional {\it learner} traits? This question is of particular interest, since a constant theme of educational data-mining over the last two decades has been an ``early warning system'' to identify students at risk in a course. 

\section{A New IRT Model for Formative Assessment}\label{sec:new}
To model guessing and copying, we introduced new traits $\gamma_j$ and $\chi_j$ into the probability of learner $j$ to solve item $i$, namely

\begin{equation}\label{eq:new2PL}
p_{ij}=\chi_j+\frac{1-\gamma_j-\chi_j}{1+\exp\left(a_i(b_i-\theta_j)\right)}\ .
\end{equation}

\begin{figure}
\begin{center}
\includegraphics[width=0.49\textwidth]{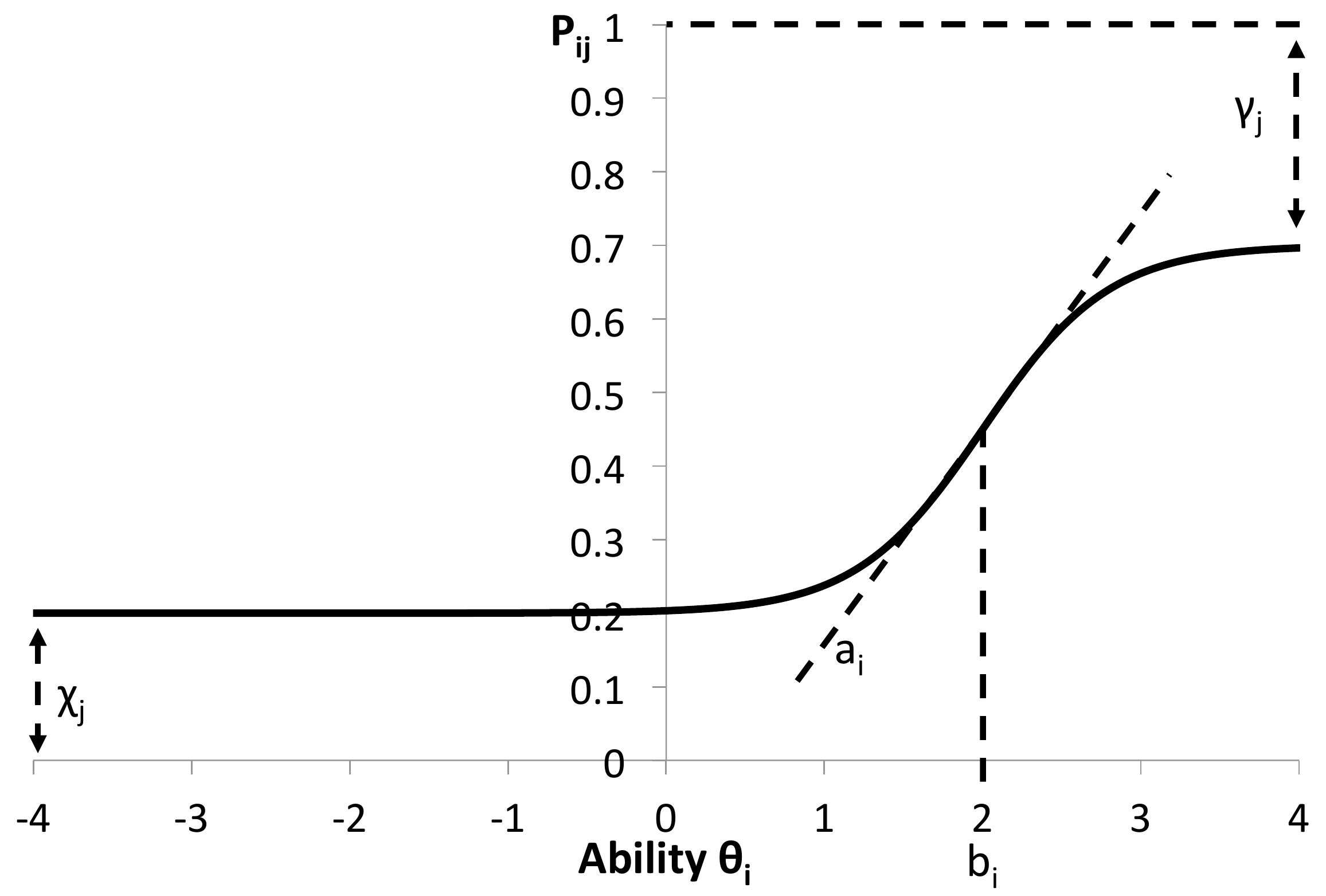}
\end{center}
\caption{Example of an item characteristic cure in the 2P3TL model. The effect of the parameters is indicated in the figure.}
\label{fig:new2PL}
\end{figure}

Fig.~\ref{fig:new2PL} illustrates this new two parameter three trait logistic model (``2P3TL'') model. The copying trait $\chi_j$ allows even low-ability students to ``solve'' problems, as it lifts the lower limit of the item characteristic curve. The guessing trait $\gamma_j$ lowers the probability of even high-ability students to get a problem correct, as they might not take the time to carefully consider the problem or verify their solution --- the trait is meant to model random guessing, i.e., inputting answers without taking the time to truly figure out the solution in spite of actually ``knowing better;'' it is different from ``last minute'' guessing due to learners being unable to arrive at a solution in spite of their best efforts --- this should continue to be modeled by the ability $\theta_j$.

The newly introduced traits $\chi_j$ and $\gamma_j$ run between zero and one; zero means that this trait is not present, while a one would indicate that a learner exhibits this behavior all the time. Another way of interpreting these parameters is that $\chi_j$ is the probability of an infinitely low-ability student to get items correct, while $\gamma_j$ is the probability of an infinitely high-ability student to fail on problems, both modeling the rates of undesirable behaviors. The trait $\theta_j$, as well as the item parameters $a_i$ and $b_i$ are in principle unrestricted, but for all practical purposes, they tend to fall into the range between negative four can four. For the purposes of the estimates performed in this study, for stability purposes, they were constrained to a range of negative to positive ten; if any of these parameters actually reach those limits, they are considered divergent.

To avoid confusion, it should be emphasized that our guessing trait $\gamma_j$ is different from the guessing  parameter $c_i$ in 3PL models~\cite{birnbaum},
\begin{equation}\label{eq:3PL}
p_{ij}=c_i+\frac{1-c_i}{1+\exp\left(a_i(b_i-\theta_j)\right)}\ .
\end{equation}
Not only is $\gamma_j$ associated with the learner (not the item), it also has the opposite effect: $c_i$ is used to model getting a 1-out-of-$N$ problem correct by merely guessing which one of the limited options could be correct, while $\gamma_j$ lowers the chances that a learner has to get an open-ended problem correct (the chances of solving a numerical or algebraic problem by merely guessing are minimal, although a really good physicist may succeed by making reasonable ``guesstimates''). Instead, in many respects, $c_i$ in a 3PL model is more similar to $\chi_j$, the copy parameter, as it improves apparent performance. In an earlier study, it was found that $c_i$ is very small; the average item parameter $c_i$ is 0.031 for first attempt homework performance, and model performance is only slightly increased by moving from 2PL to 3PL~\cite{kortemeyer14a} --- but once again, that was attempting to model copying as a property of a particular item, not a particular learner; in other words, it assumed that some problems are more likely to be copied than others (which was not the case~\cite{kortemeyer14a}), rather than that some learners copy more frequently than others.

\section{Results}\label{sec:results}
The new model was explored using data from the same physics course already investigated in Ref.~\cite{kortemeyer14a}.  First, we investigated how well the learner ability estimated from homework reflects the ``true'' ability estimated from the exam data. In a second step we investigated the model influence on item parameters. To better understand the characteristics of the new traits, in both of these studies, they were introduced one-at-a-time. Finally, we compare the newly introduced traits to self-reported gender-dependent data on homework behavior.

\subsection{Learner Ability}
As in Fig.~\ref{fig:firstlastnone}, the data from the exams was modeled using a standard 2PL model (Eq.~\ref{eq:2PL}) and assumed to be a true reflection of the learners' abilities.

\subsubsection{Correlation of a Reduced ``Guessing Only'' Model}
\begin{figure*}
\begin{center}
\includegraphics[width=0.49\textwidth]{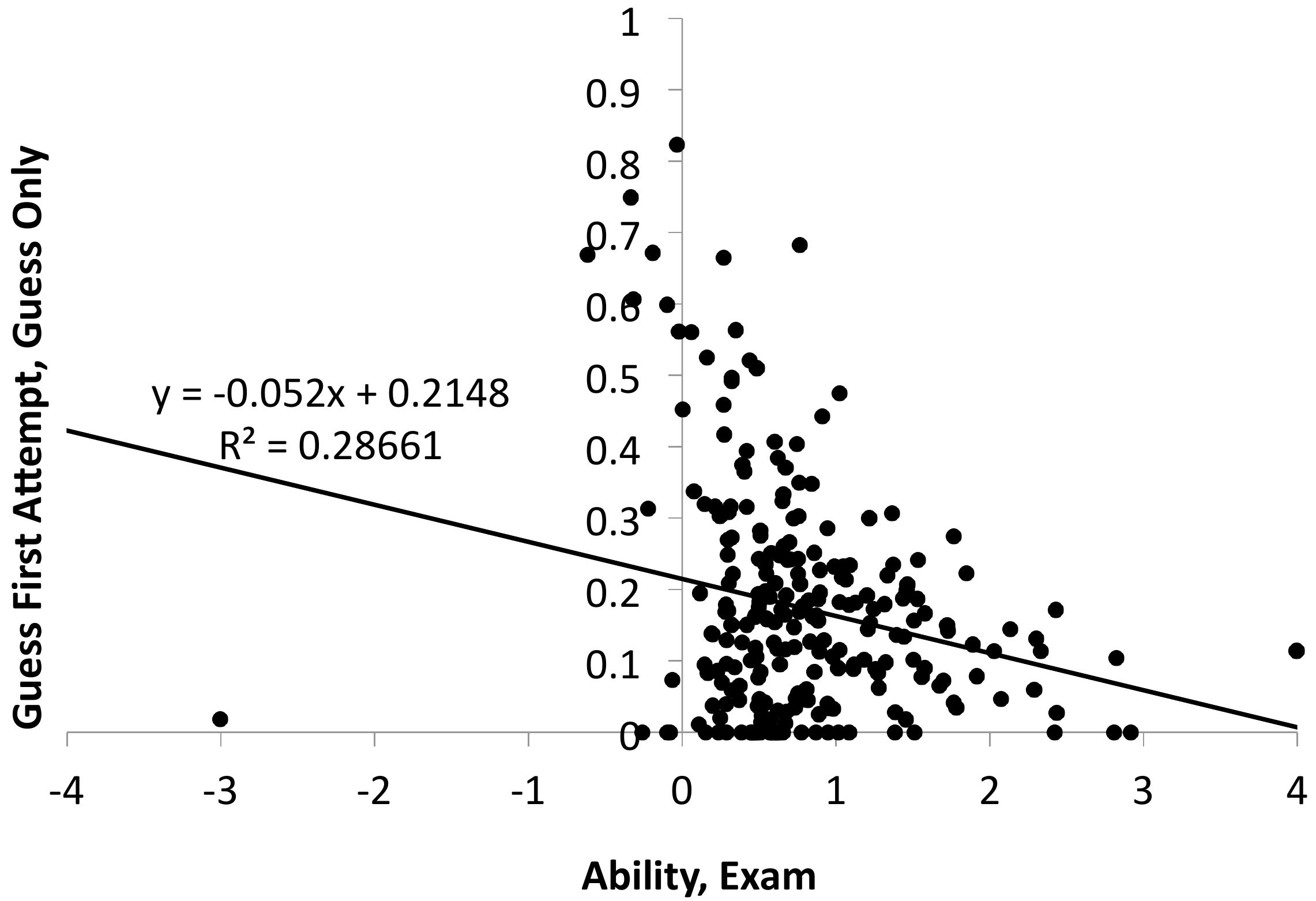}
\includegraphics[width=0.49\textwidth]{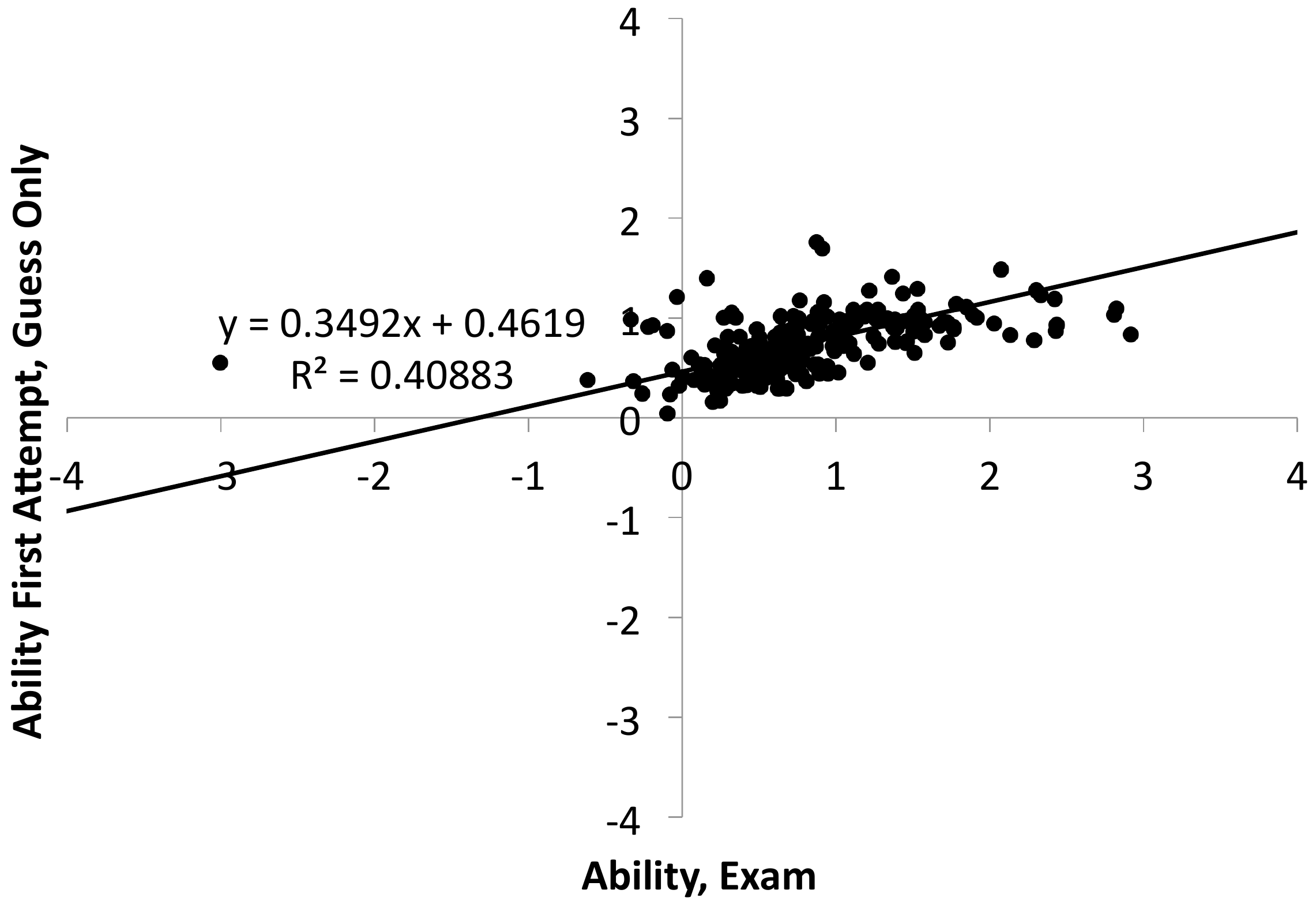}
\end{center}
\caption{Learner guessing trait $\gamma_j$ (left panel) and ability $\theta_j$ (right panel) based on a reduced 2P3TL model ($\chi_j\equiv0$) of the first-attempt homework data versus ability based on exam data.}
\label{fig:firstguess}
\end{figure*}

Fig.~\ref{fig:firstguess} shows the outcome of a reduced 2P3TL model for the first-attempt homework data, where the copying trait $\chi_j$ was suppressed ($\chi_j\equiv0$), in correlation to the ability derived from the exam data. It is apparent that low-ability learners guess more than high-ability learners (Fig.~\ref{fig:firstguess}, left panel), which is not surprising: high ability on exams may well be correlated with more diligence while doing homework. On their first attempt, some students guess (wrong) up to eighty percent of the time, but zero to twenty percent is more typical. This is a known phenomenon: when a lot of attempts are offered to get a problem correct, some learners tend to be careless about wasting attempts~\cite{kortemeyer15a}.

Compared to the left panel of Fig.~\ref{fig:firstlastnone}, the model tends to assign higher abilities to learners (Fig.~\ref{fig:firstguess}, right panel), since in this model, failure to succeed on problems is not necessarily due to low ability; instead, by design, some percentage of failure can get attributed to random guessing in spite of actually knowing better. The correlation between homework and exam ability, however, did not improve: the model has more variability, but explains slightly less of it.
\subsubsection{Correlation of a Reduced ``Copying Only'' Model}
\begin{figure*}
\begin{center}
\includegraphics[width=0.49\textwidth]{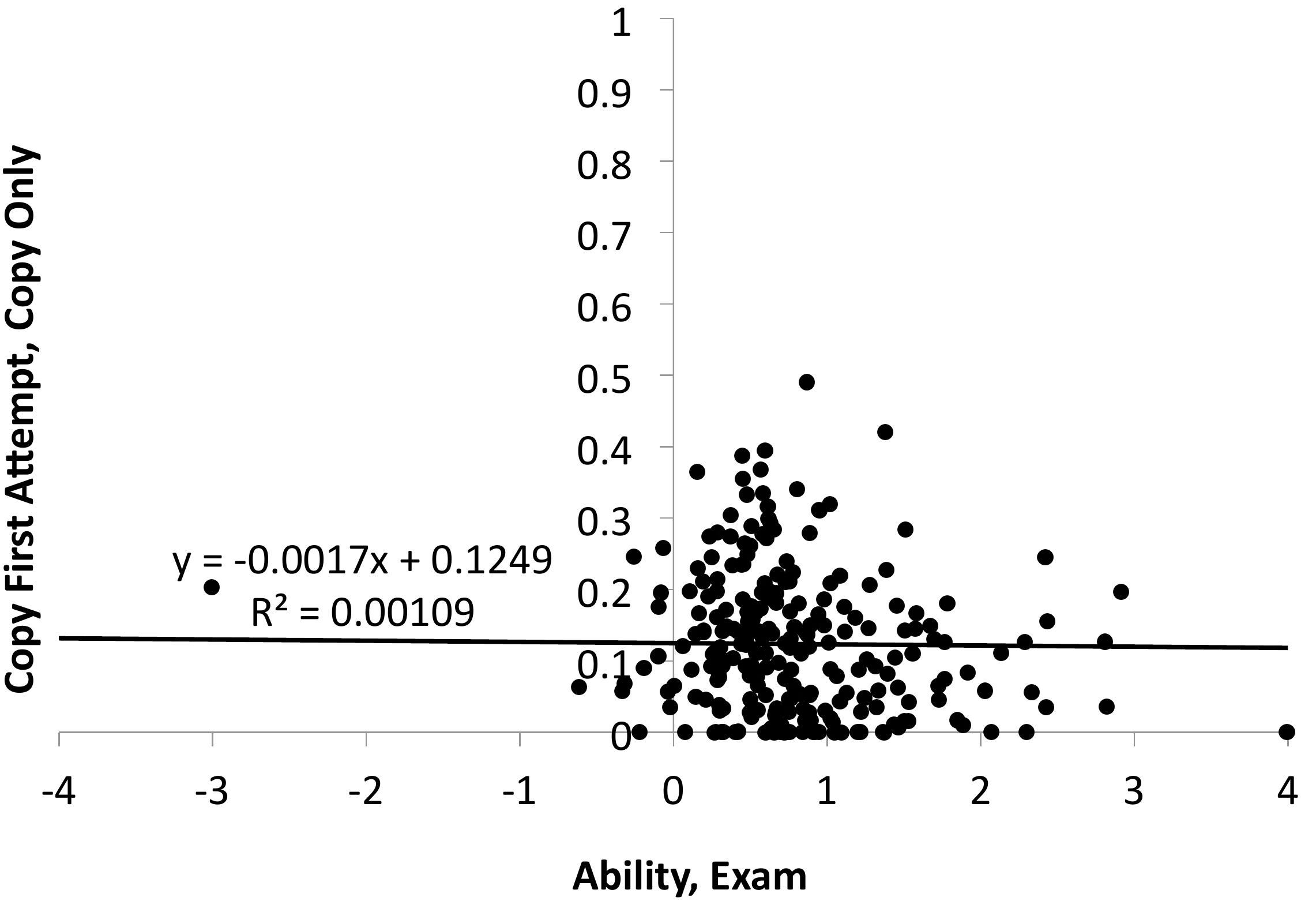}
\includegraphics[width=0.49\textwidth]{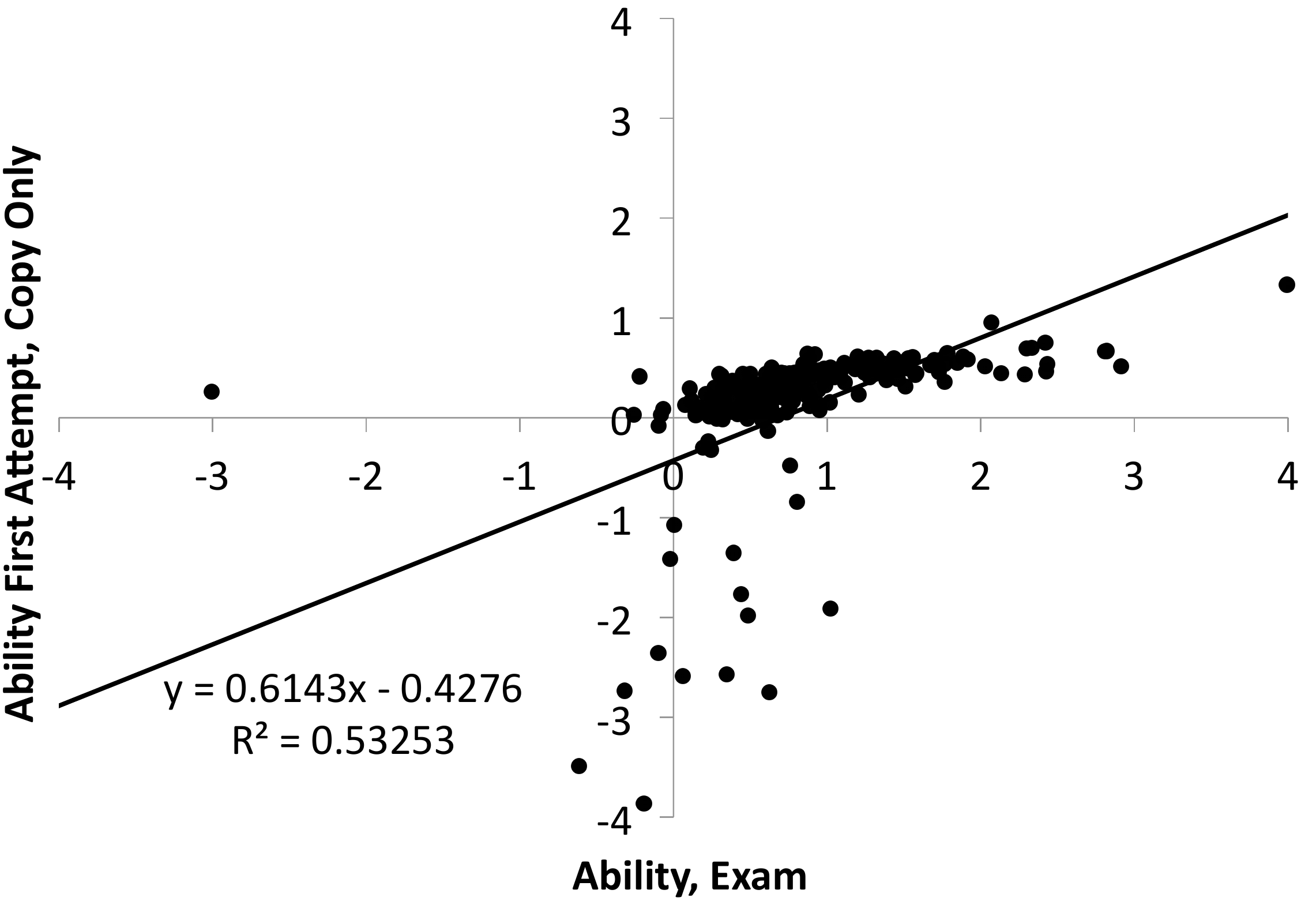}
\end{center}
\caption{Learner copying trait $\chi_j$ (left panel) and ability $\theta_j$ (right panel) based on a reduced 2P3TL model ($\gamma_j\equiv0$) of the first-attempt homework data versus ability based on exam data.}
\label{fig:firstcopy}
\end{figure*}

Fig.~\ref{fig:firstcopy} shows the outcome of a reduced 2P3TL model for the first-attempt homework data, where the guessing trait $\gamma_j$ was suppressed ($\gamma_j\equiv0$).  It is interesting to note that in this model, copying does not appear to depend on ability: all learners copy, on the average about twelve percent of the time. There are some notable outliers, though, who according to this model copy half of their first homework attempts from others.

Not surprisingly, the estimated ability based on homework of some students decreased, most notably those with abilities lower than unity on the exam-based ability scale. While, as the right panel shows, these students do not necessarily copy more, their first-attempt successes are apparently mostly due to copying. Also not surprisingly, the predictive power of student ability improved by absorbing copying into another trait, since at least in theory, students are not able to copy on exams. In fact, remarkably, the predictive power is now better than the estimates based on the 2PL analysis of the final homework attempt (Fig.~\ref{fig:firstlastnone}, right panel).
\subsubsection{Correlation of the Full Model}
\begin{figure*}
\begin{center}
\includegraphics[width=0.49\textwidth]{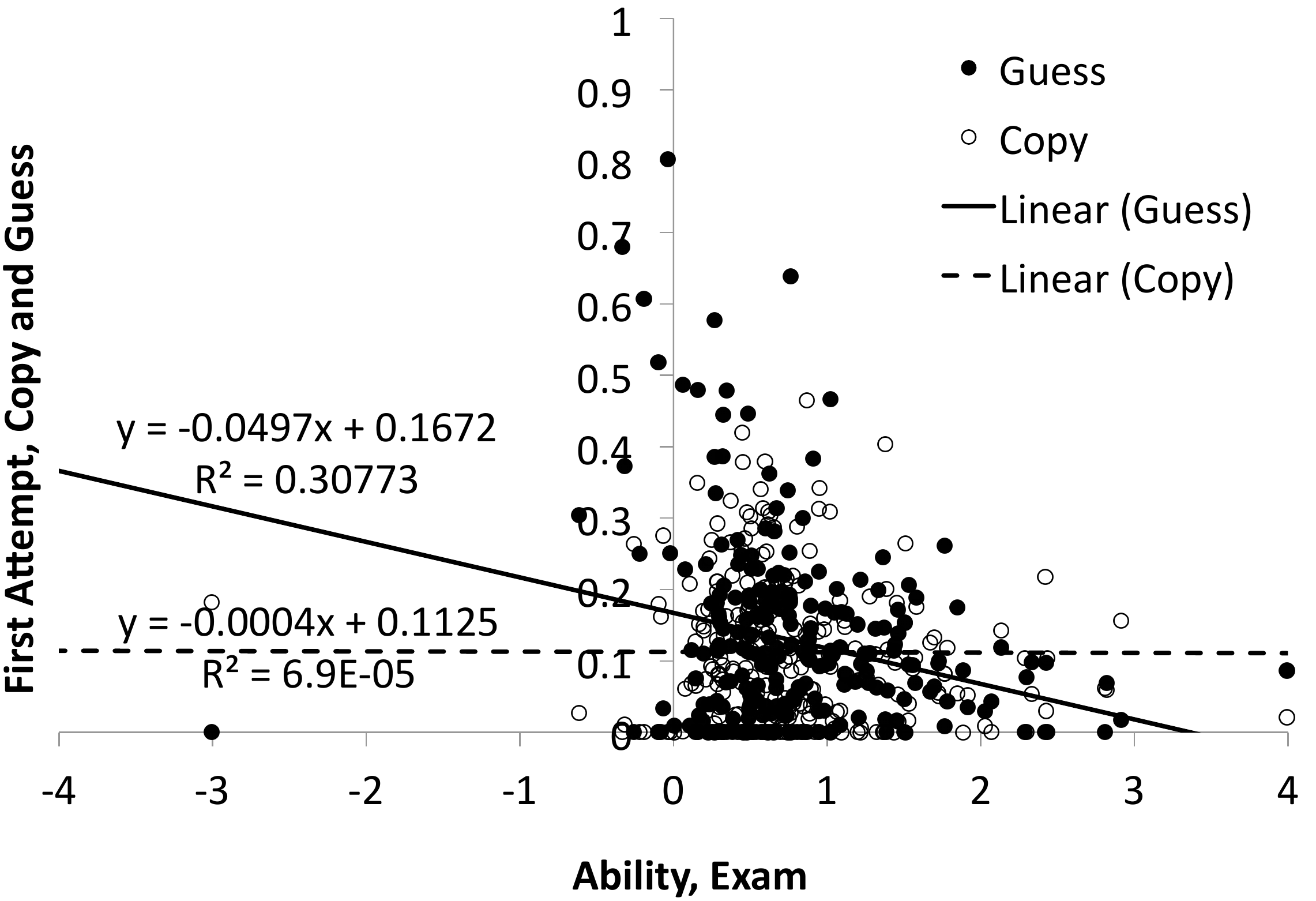}
\includegraphics[width=0.49\textwidth]{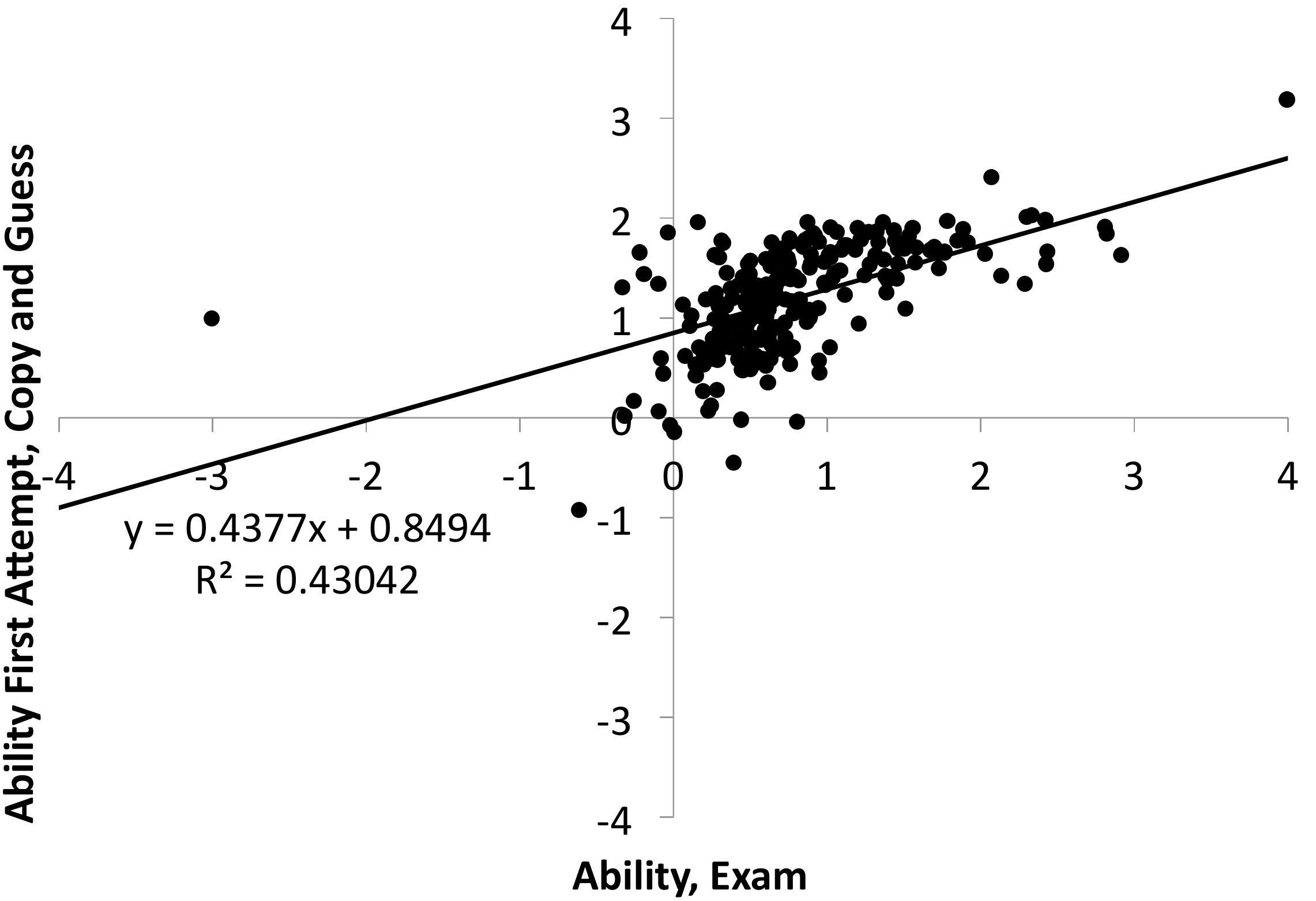}
\end{center}
\caption{Learner guessing and copying traits $\gamma_j$ and $\chi_j$ (left panel) and ability $\theta_j$ (right panel) based on the full 2P3TL model of the first-attempt homework data versus ability based on exam data.}
\label{fig:firstall}
\end{figure*}

Fig.~\ref{fig:firstall} shows the outcome of the full 2P3TL model for the first-attempt homework data. Somewhat counter-intuitively, the model predictive power decreases. The correlations between the guessing and copying traits in this full model on the one hand and exam-ability on the other are almost the same as in the respective reduced models (comparing the right panel of Fig.~\ref{fig:firstall} to the right panels of Figs.~\ref{fig:firstguess} and~\ref{fig:firstcopy}), suggesting that these two traits are independent (which we hoped for). However, the predictive power for the ability somewhat decreased compared to the reduced model with only the copying trait, suggesting that with the two new traits, not enough homework variability is left to estimate the full ability spectrum. While still better than the 2PL ability derived from the final attempt, the full 2P3TL model might be overfitting the data.

\subsubsection{Final Instead of First Homework Attempt}
Applying the 2P3TL model to the data for the final attempt did not improve the model predictive power; in fact, the ability estimates for a handful of very high and very low performing students started to diverge. A learner's final attempt is either whatever attempt they first succeeded solving the problem, or the last attempt before they abandoned the problem or ran out of tries. Somewhat disappointingly, it was found earlier that subsequent attempts are independent of each other (i.e., students do not appear to learn from earlier failed attempts)~\cite{kortemeyer15a}, but it is unlikely that the probability of guessing and copying is independent of the number of attempts. If students are copying, they likely do so from the start (i.e., on the first attempt), and they are more likely to ``just guess'' when there are a lot of attempts left. It appears that the complexity introduced by these behavior patterns could not be modeled well by the new traits.

\subsubsection{Distributions}
\begin{figure*}
\begin{center}
\includegraphics[width=0.98\textwidth]{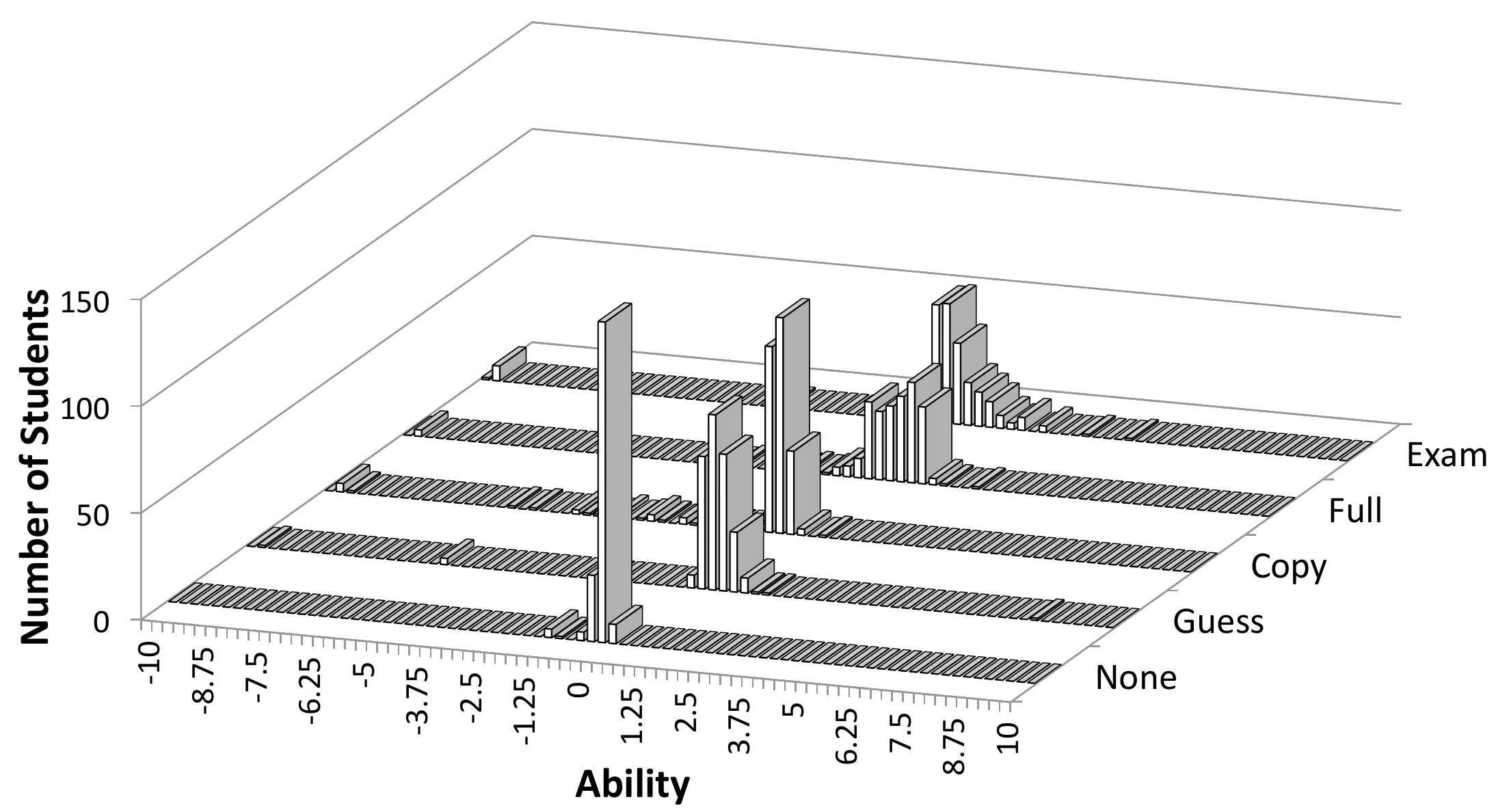}
\end{center}
\caption{Distribution of learner ability $\theta_j$ in a 2PL model of the homework data (first row), as well as ``guessing only, ``copying only'' and full 2P3TL models (second through fourth row, respectively). For comparison, the distribution of the learner ability estimated from exam data is given in the fifth row.}
\label{fig:studentabi}
\end{figure*}

Fig.~\ref{fig:studentabi} shows the distribution of the learner ability $\theta_j$ in a 2PL model of the homework data, as well as ``guessing only, ``copying only'' and full 2P3TL models, compared to the estimation based on the exam data. With the introduction of new learner traits, the distribution of estimated learner ability widens; in fact, the bulk of the learner ability distribution of the full model is comparable to that of the exam data. 

However, even though the distributions look similar between the full 2P3TL model of the homework and the 2PL model of the exam, it was found earlier that the ``copying only'' estimates correlate better. This somewhat surprising outcome is largely due to the approximately 6\% of the students for which the ability estimate dramatically decreased and in fact became negative (see Fig.~\ref{fig:firstcopy}) --- separating these students (who also have low exam-based ability) from the bulk of the distribution increased the correlation. Introducing ``guessing'' in the full model pulled these students pack into the bulk of the distribution: the same students who copy also guess (guessing increases the estimate of the true ability).

\subsection{Item Parameters}
For the item parameters, no comparison to exam data is possible, as there was no direct overlap between items used in homework and those used on exams. 

\subsubsection{Difficulty Distribution}
\begin{figure*}
\begin{center}
\includegraphics[width=0.98\textwidth]{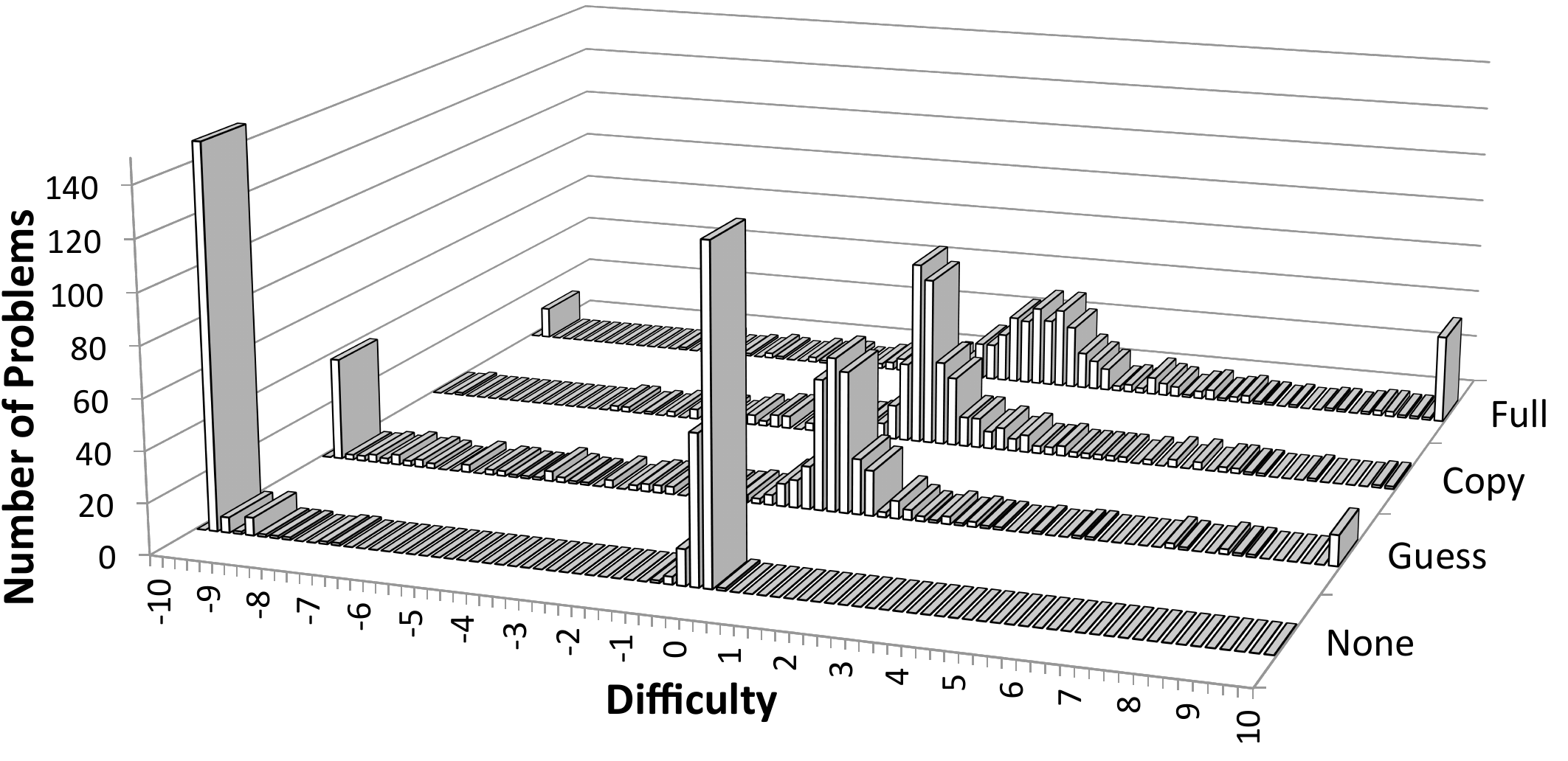}
\end{center}
\caption{Distribution of the difficulty parameter $b_i$ in a 2PL model of the homework data (first row), as well as ``guessing only, ``copying only'' and full 2P3TL models (second through fourth row, respectively).}\label{fig:itemdiff}
\end{figure*}

Fig.~\ref{fig:itemdiff} shows the distribution of the estimated difficulty parameters for the homework items. The failure of the 2PL is readily evident from the large number of divergent estimates (as noted earlier, boundaries of from negative to positive ten were imposed on the difficulty), a problem that was already identified in the earlier study~\cite{kortemeyer14a}. Introducing the additional student traits stabilizes the model, where in fact the reduced ``copying only'' model completely absorbs any divergencies.
Items are shifted toward higher difficulty, which is not surprising: the copying trait explains why low-ability students might get those items correct and thus moves up the difficulty, which is based on the true ability of the learner.

\subsubsection{Discrimination Distribution}
Fig.~\ref{fig:itemdiff} shows the distribution of the estimated discrimination parameters for the homework items. As already found earlier~\cite{kortemeyer14a}, the discrimination parameter is more prone to divergence than the difficulty, mostly due to the fact that with increasing  $a_i$ the item characteristic curve Fig.~\ref{fig:new2PL} becomes so steep at $b_i$ that further changes to $a_i$ make little difference in the quality of the fit --- if no artificial constraints are introduced, the parameter starts to drift. While the introduction of the additional learner traits does not eliminate this problem for some items, the bigger difference occurs for items with little or no discrimination, i.e., the large peak at $a_i=0$ in the distribution of the 2PL model. The additional learner traits absorb some of the noise that drowns out the item discrimination, and a larger number of items now show positive discrimination.

\begin{figure*}
\begin{center}
\includegraphics[width=0.98\textwidth]{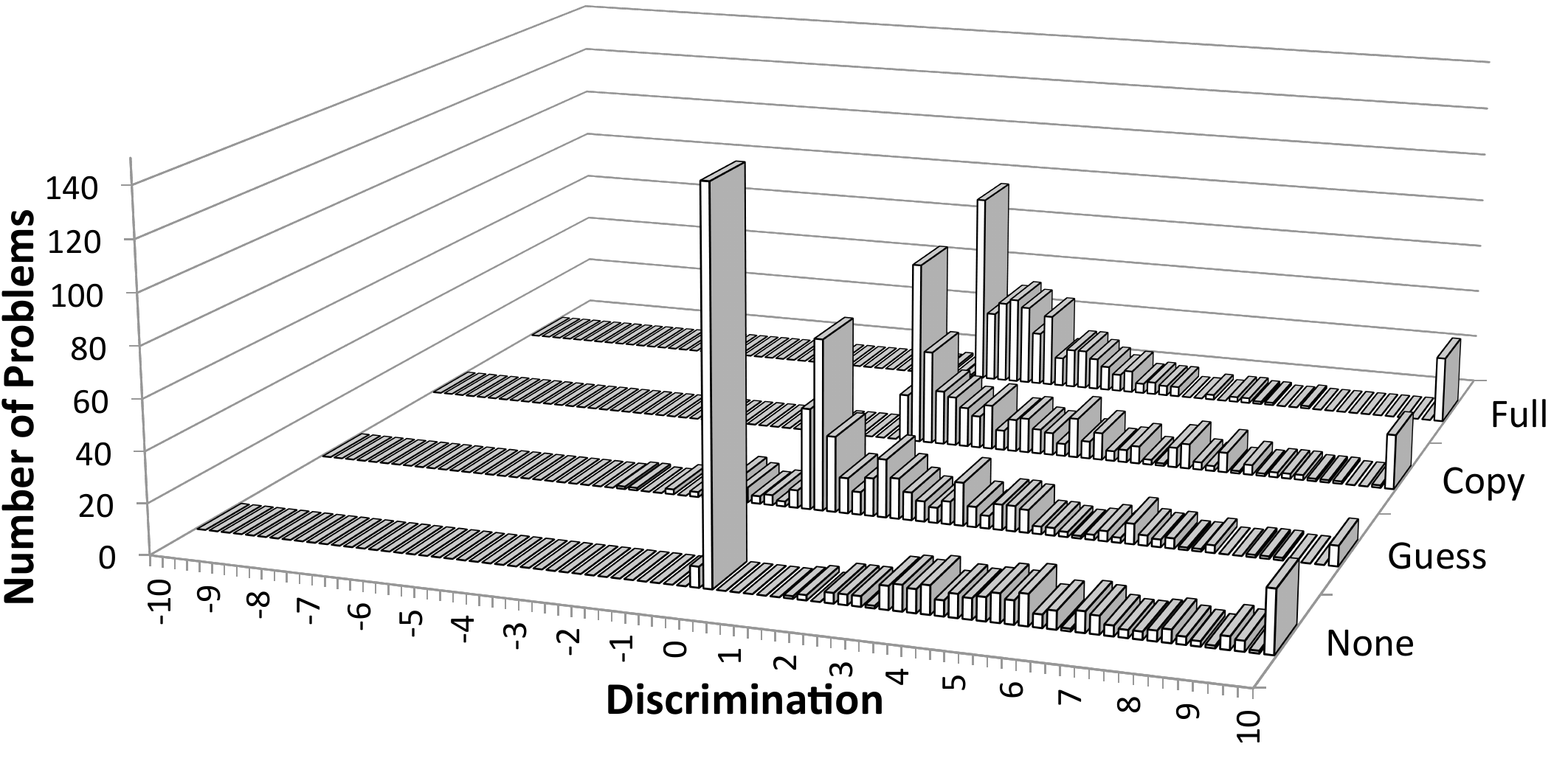}
\end{center}
\caption{Distribution of the discrimination parameter $a_i$ in a 2PL model of the homework data (first row), as well as ``guessing only, ``copying only'' and full 2P3TL models (second through fourth row, respectively).}\label{fig:itemdisc}
\end{figure*}
\subsection{Relationship with Self-Reported Homework Behavior}
In an earlier study it was found that male and female students interact differently with online homework~\cite{kortemeyer09}. When asked about their first action on a new homework problem, 58\% of the male students answered that the ``immediately attempt'' the problem, while only 39\% of the female students stated the same; in addition, 14\% of the male students stated that they ``submit random stuff or guess,'' while only 8\% of the female students did so; it is to be expected that this behavior should be reflected in the guessing trait $\gamma_j$.

On the other hand, only 8\% of the male students stated that the first thing they do with a new problem is discuss it with teaching assistant, friends, or in the online discussions, while 20\% of the female students do so. This kind of social interaction or collaboration can take many forms, but it might be reflected in the copying trait $\chi_j$.

\begin{figure*}
\begin{center}
\includegraphics[width=0.49\textwidth]{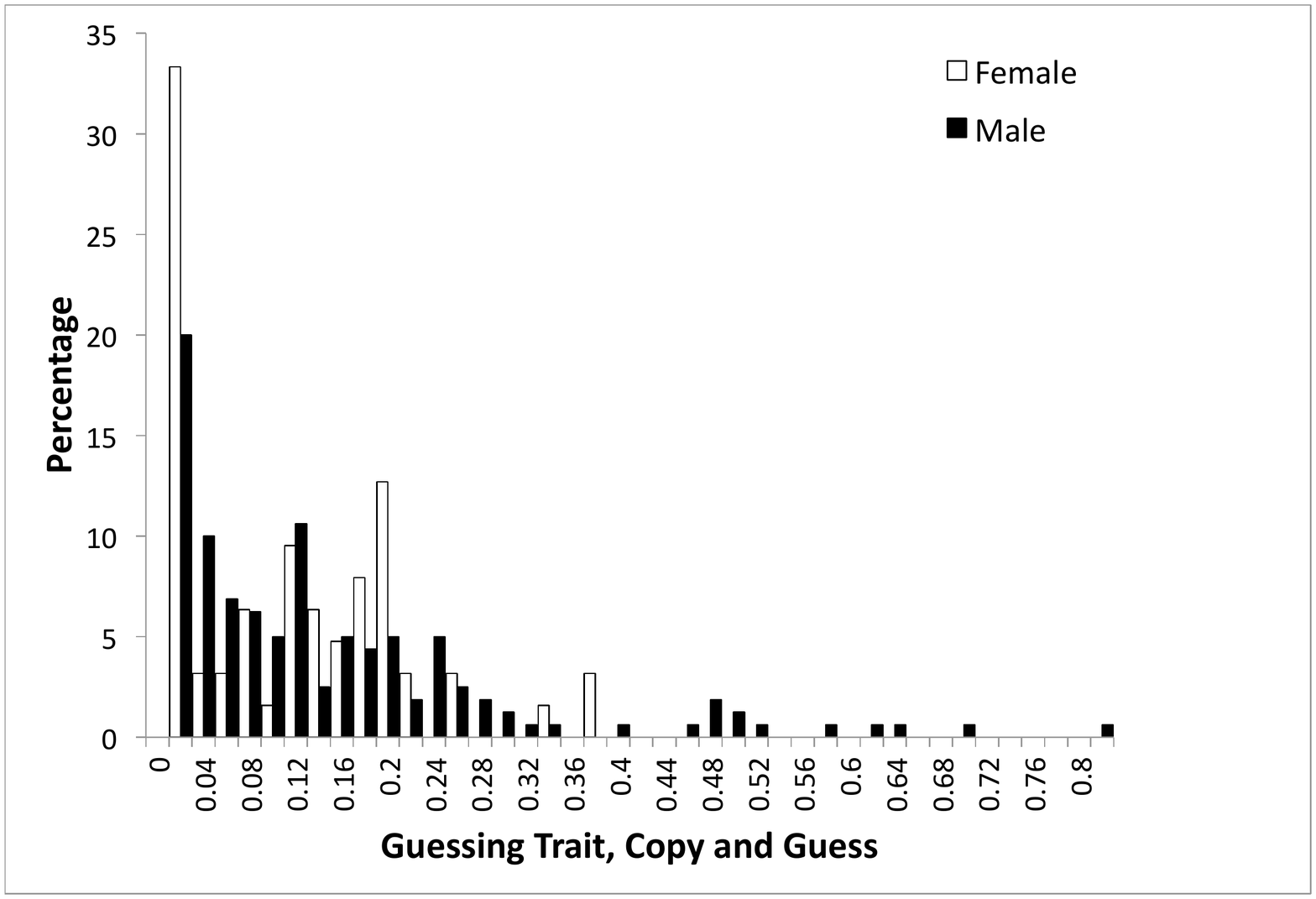}
\includegraphics[width=0.49\textwidth]{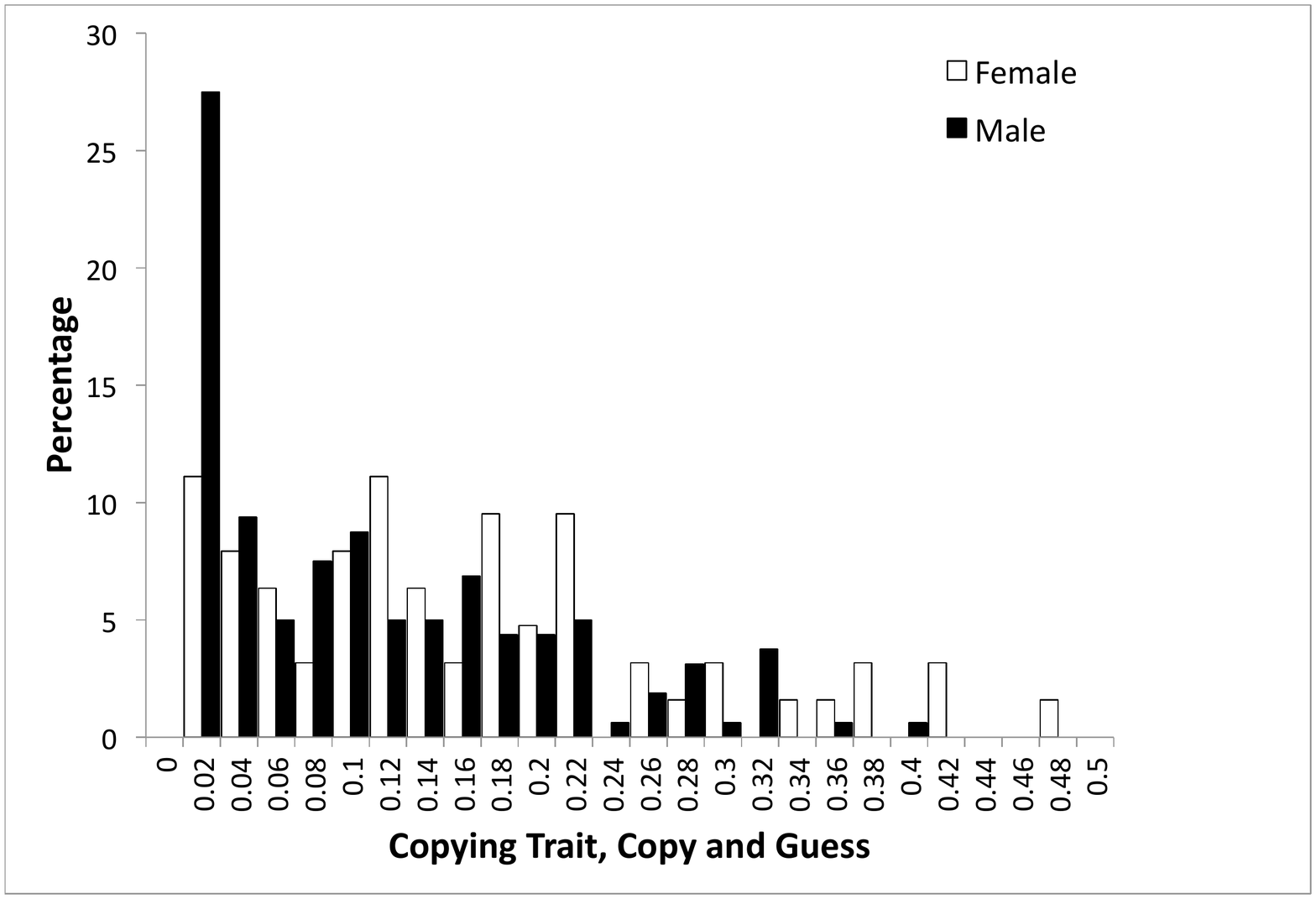}
\end{center}
\caption{Histograms of the learner guessing and copying traits $\gamma_j$ (left panel) and $\chi_j$ (right panel) based on the full 2P3TL model of the first-attempt homework data, separated by gender.}
\label{fig:gender}
\end{figure*}

Fig.~\ref{fig:gender} shows histograms of the guessing and copying traits, separated by gender, for the full 2P3TL model. It is apparent that indeed male students tend to guess more than female students ($\gamma_{\mbox{\scriptsize ave}}=0.16\pm0.20$ for male versus  $\gamma_{\mbox{\scriptsize ave}}=0.10\pm0.10$ for female students), and they copy less ($\chi_{\mbox{\scriptsize ave}}=0.10\pm0.09$ for male versus  $\chi_{\mbox{\scriptsize ave}}=0.15\pm0.11$ for female students), though neither difference is statistically significant. Large differences occur in the distributions at $\gamma_j\approx0$ and $\chi_j\approx0$, as well as at the tail ends. The ability trait, in the other hand, comes out almost equal ($\theta_{\mbox{\scriptsize ave}}=1.07\pm1.40$ for male versus  $\theta_{\mbox{\scriptsize ave}}=0.96\pm0.46$ for female students). As gender was not part of the IRT estimate, it is encouraging that the newly introduced traits nevertheless independently correspond to self-reported data.

\section{Discussion}\label{sec:disc}
Under the assumption that the newly introduced item parameters $\gamma_i$ and $\chi_i$ indeed model guessing and copying, it is apparent that for the first attempt on online homework, copying is distorting reliable performance feedback more strongly than randomly guessing. Learners are apparently less prone to begin work on a problem by inputting random answers than they are to copy right away; guessing may appear in later attempts out of desperation, but as discussed, this is a sign of low ability. In fact, the first attempt on a homework problem item may best be modeled by
\begin{equation}\label{eq:newRed2PL}
p_{ij}=\chi_j+\frac{1-\chi_j}{1+\exp\left(a_i(b_i-\theta_j)\right)}\ ,
\end{equation}
without the guessing parameter $\gamma_j$. The fact that the trait $\chi_j$ can be rather large (up to 0.5 for one of the learners), and that it is around 0.125 on the average, compared to an average of 0.031 (Ref.~\cite{kortemeyer14a}) for the ``copying'' parameter $c_i$ in the otherwise mathematically similar Eq.~\ref{eq:3PL}, confirms the assumption that copying behavior is tied to the learner, not the item.

The result suggests that getting an online homework problem correct on the first attempt might be a very deceptive measure of learner ability; instead, immediate success might simply be the result of cheating. Truly high-ability students might take multiple attempts as they learn the material. Decreasing credit for homework based on the number of attempts used to get the correct result might aggravate this situation and thus be counter-productive, as copying gets rewarded.

\section{Outlook}\label{sec:outlook}
While copying appears to be the dominant learner trait during the first attempt on a homework problem, in a more extensive study, it could be investigated if there is a dependence on the number of subsequent attempts (e.g., more copying or random guessing after a few failed attempts), and if for example copying or random guessing occur more frequently as the learner is about to run out of attempts or as the deadline approaches.

The learner traits extrapolated from this model can be used as an early-warning system for undesirable student behavior and learners-at-risk. IRT parameter estimation is essentially a multidimensional optimization problem, and a challenge with the implementation of the new model in online homework systems is scale: the estimations for this study of 256 students and 401 homework items required approximately five minutes of computation time on a modern workstation (2014 hardware), using a simplified Markov Chain Monte Carlo (MCMC) algorithm (based on Ref.~\cite{patz99}) implemented in FORTRAN. Unfortunately, the runtime for the algorithm scales like
\begin{equation}
T_{\mbox{\scriptsize run}}\sim(N_{\mbox{\scriptsize parameters}}+N_{\mbox{\scriptsize traits}})N_{\mbox{\scriptsize students}}N_{\mbox{\scriptsize items}}\ ,
\end{equation}
i.e., proportional to the number of items and the number of students (as well as the sum of the number of parameters and traits). In a large homework systems, where the same items may be used across a number of classes, and the same students may take more than one class in the same or subsequent semesters, the number of items and students can easily be three orders of magnitude larger, which would result in almost a decade of computation time. Thus, mechanisms need to be found to iteratively update the estimates as transactions occur~\cite{kortemeyer15b}.

The new model can also be used for additional physics education research projects, as it unearths learner traits that otherwise could only be found through observation or self-reporting; both of those methods may influence the results, as students may be less likely to exhibit or report undesirable homework behavior when observed or surveyed. In contrast, the presented model can extract these behaviors from the raw homework data. It will be interesting to correlate these results with other student characteristics (beyond gender) to learn more about the dynamics of online homework. 

\section{Conclusions}\label{sec:conclusions}
By introducing two new learner traits into an IRT model, we were able to increase the predictive power of the learner ability derived from the first attempt on solving online homework problems. These two parameters model copying and random guessing behavior, and it was shown that absorbing these undesirable behaviors in those two traits moves the estimated ability closer to the ``true'' ability estimated from exam data. It was found that on the average about 12\% of the initial homework submissions might be copied, almost independent of learner ability. Random guessing contributes less to the noise that is masking the true learner ability, and it was found that low-ability students guess more than high-ability students. The introduced learner traits also improve the estimates of the traditional item parameters difficulty and discrimination: the convergence of the difficulty parameter improves greatly, and more items exhibit higher positive discrimination. Given the relative impact of the new traits, future studies might limit themselves to the copying trait only.
\bibliographystyle{apsper}
\bibliography{bibfile}

\end{document}